\documentclass{article}
\usepackage{ijcai23}

\usepackage{times}
\usepackage{soul}
\usepackage{url}
\usepackage[hidelinks]{hyperref}
\usepackage[utf8]{inputenc}
\usepackage[small]{caption}
\usepackage{graphicx}
\usepackage{amsmath}
\usepackage{amsthm}
\usepackage{booktabs}
\usepackage{algorithm}
\usepackage{algorithmic}
\usepackage[switch]{lineno}

\usepackage{multicol, multirow}
\usepackage{threeparttable}
\usepackage{bbding}
\usepackage{pifont}
\usepackage{xcolor}
\usepackage{wasysym}
\usepackage{amssymb}

\title{Feature Representation Learning for Click-through Rate Prediction: A Review and New Perspectives}

\iftrue
\author{
Fuyuan Lyu$^1$\and
Xing Tang$^2$\and
Dugang Liu$^{3}$\and
Haolun Wu$^{1,4}$\and \\
Chen Ma$^5$\and
Xiuqiang He$^2$ \And
Xue Liu$^1$
\affiliations
$^1$School of Computer Science, McGill University\\
$^2$FiT, Tencent\\
$^3$Guangdong Laboratory of Artificial Intelligence and Digital Economy (SZ)\\
$^4$Mila - Quebec Artificial Intelligence Institute\\
$^5$Department of Computer Science, City University of Hong Kong
\emails
\{fuyuan.lyu, haolun.wu\}@mail.mcgill.ca,
\{shawntang,xiuqianghe\}@tencent.com,
dugang.ldg@gmail.com,
chenma@cityu.edu.hk,
xueliu@cs.mcgill.ca
}
\fi

\begin{document}

\maketitle

\begin{abstract}
Representation learning has been a critical topic in machine learning. In Click-through Rate Prediction, most features are represented as embedding vectors and learned simultaneously with other parameters in the model. With the development of CTR models, feature representation learning has become a trending topic and has been extensively studied by both industrial and academic researchers in recent years. This survey aims at summarizing the feature representation learning in a broader picture and pave the way for future research. To achieve such a goal, we first present a taxonomy of current research methods on feature representation learning following two main issues: (i) which feature to represent and (ii) how to represent these features. Then we give a detailed description of each method regarding these two issues. Finally, the review concludes with a discussion on the future directions of this field.
\end{abstract}

\section{Introduction}
\label{sec:intro}

The success of machine learning largely depends on data representation~\cite{RL}. As a machine learning application in real-world, Click-through rate (CTR) prediction plays as a core function module in various personalized online services, including online advertising, recommender systems, web search, to name a few.~\cite{DLsurvey}. 

\begin{figure}[!htbp]
    \centering
    \includegraphics[width=0.5\textwidth]{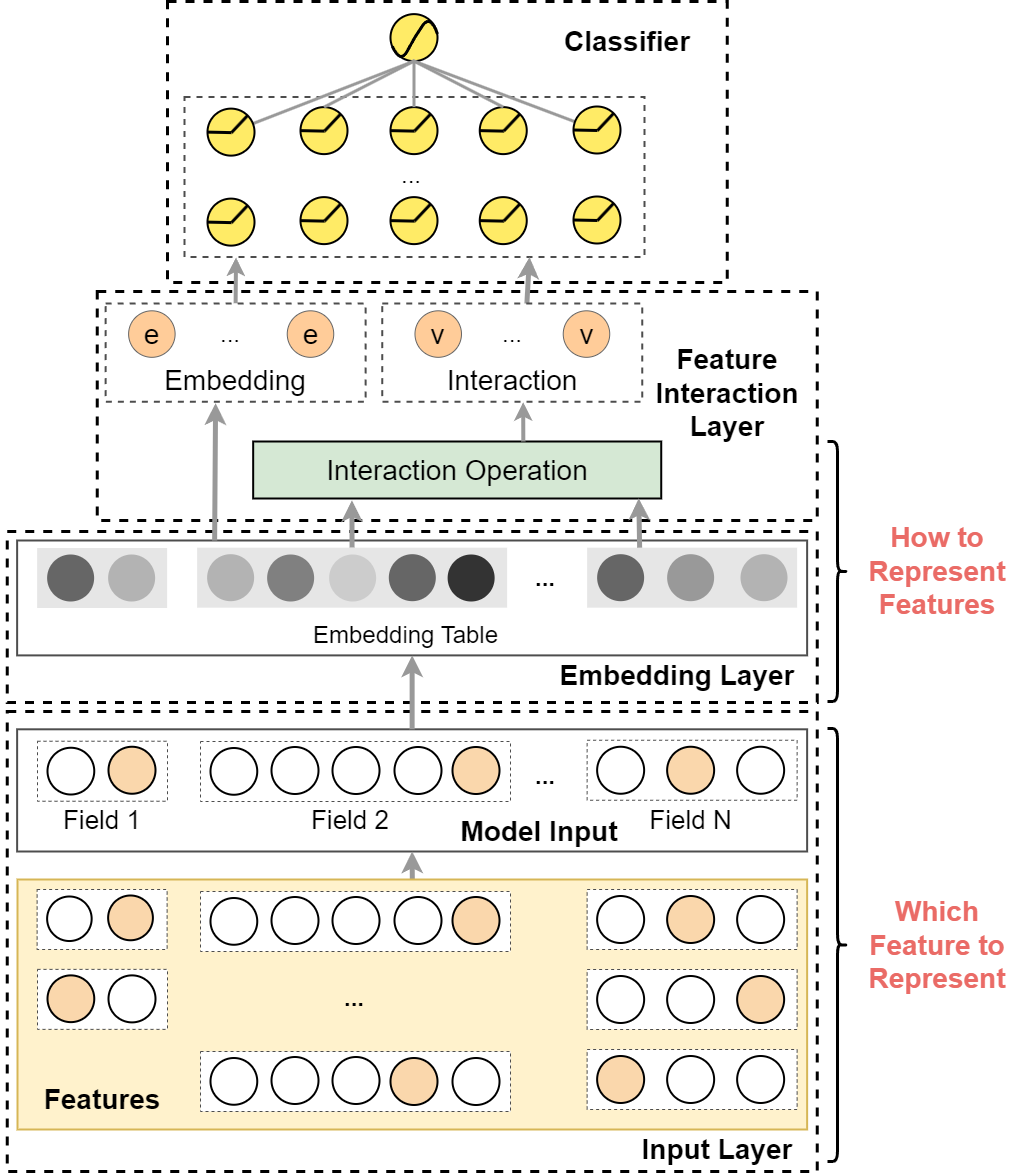}
    \caption{Overview of the general framework of CTR prediction.}
    \label{fig:ctr}
\end{figure}

Generally speaking, the typical inputs of CTR models consist of many categorical features. We term the values of these categorical features as feature values, which are organized as feature fields. For example, a feature field \textit{gender} contains three feature values, \textit{male}, \textit{female} and \textit{unknown}. These predictive models use the embedding table to map the categorical feature values into real-valued dense vectors. Then these embeddings are fed into the feature interaction layer, such as factorization machine~\cite{FM}, cross network~\cite{DCN}, and Squeeze-and-Excitation layer~\cite{FiBiNet}. The final classifier aggregates the representation vector to make the prediction. The whole pipeline is shown in Figure \ref{fig:ctr}. Hence, the performance of these CTR models is heavily dependent on the choice of feature representation on which they are applied~\cite{RL}.

How to learn feature representation for CTR has been investigated since factorization machine~\cite{FM}, which introduce latent vector to model features. However, there are still a lot of work have focused on certain issues of feature representation learning for CTR in both academia and industry recently \cite{AutoField,OptEmbed,AutoFIS}. Moreover, there are also some related survey recently. ~\cite{DLsurvey} reviews the development of deep CTR model while ~\cite{AutoMLsurvey} focuses on the application of automated machine learning in recommender system. In this paper, we summarize the representation learning for the CTR prediction according to two important issues for the first time: i) which feature should be represented in the feature representation learning and ii) how we should represent these features. Hence, our discussion shed some light on this problem and provide some new perspectives for feature representation in CTR domain.

Based on the two issues aforementioned, we organize the rest of the paper as follows: First, we provide a rough taxonomy of previous work with respect to these two issues in Section \ref{sec:taxonomy}. A unified formulation of feature representation learning is presented in Section \ref{sec:form}. We also detailedly illustrate previous works that address both issues in Section \ref{sec:which} and Section \ref{sec:how}, respectively. Finally, the review concludes with several potential directions of the feature representation in Section \ref{sec:summary}.

\begin{figure*}[!htbp]
    \centering
    \includegraphics[width=0.95\textwidth]{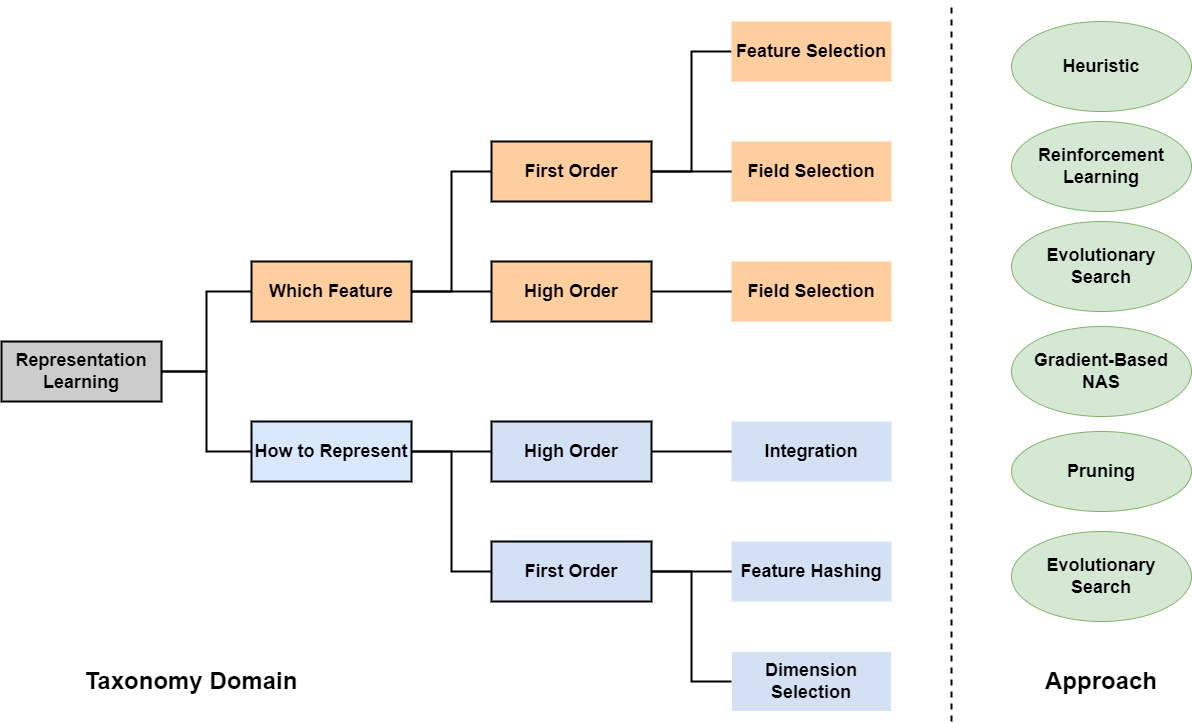}
    \caption{Taxonomy of feature representation learning and related approaches.}
    \label{fig:taxonomy}
\end{figure*}

\section{Taxonomy}
\label{sec:taxonomy}

We differentiate representation learning in the CTR prediction task by which feature should be represented and and how we should represent them. An overview of the taxonomy of the feature representation learning and the approaches adopted to deal with these two issues is outlined in Figure \ref{fig:taxonomy}.

Considering the first one: \textbf{which feature to represent}, some studies focus on the selection of first order features (a.k.a. original features), as feature engineering proves to be essential towards accurate prediction. These methods can be further categorized by their granularity of the selection: field or feature. Beyond the selection of first order features, other researchers focuses on selecting second or high-order features. However, only field-level selection methods are designed for second or high-order features due to combination explosion of search space.

Another important issue for feature representation learning in CTR is \textbf{how to represent these selected features}. 
For first order features, one of the major differences lies in the choice of embedding dimensions. Higher dimensions may lead to overfitting representation, while lower dimensions may lead to underfitting. 
Another important topic regarding first order features is about feature hashing, which maps multiple features into a same representation. The design of feature hashing is practically related to the large-scale industrial models with hardware requirements.
For higher-order features, the major work relies on how to generate their representations. For instance, most works~\cite{AutoFeature} represent the features as the output of an aggregation function which takes related first order feature representations as input. Therefore, the selection of such aggregation functions becomes an essential component. 

Various approaches have also been adopted by different feature representation learning methods to either solve these two factors. These approaches can mainly be categorized into many classes: Reinforcement Learning-based~\cite{NAO}, Statistics-based~\cite{PCA,LASSO}, Edge Prediction, Gradient-based Neural Architecture Search(NAS)~\cite{DARTS}, Network Pruning~\cite{STR} and Evolutionary Search-based NAS.

\section{Formulation}
\label{sec:form}

In our setting, we represent all possible first-order features as $\mathbf{X}^{(1)} = \{\mathbf{x}_{1}, \mathbf{x}_{2}, \cdots, \mathbf{x}_{m}\}$. Here $\mathbf{x}_{i}$ is a one-hot encoded representation, which tends to be sparse and high-dimensional. Whereas each field $\mathbf{z}_i$ contains a proportion of all possible first-order features, denoted as:
\begin{equation}
    \mathbf{z}_i = \{ \mathbf{x}_{i_p} \}, \ 1 \le i_p \le m,
\end{equation}
For instance, the \textit{gender} field $\mathbf{z}_1$ contains three features $\mathbf{z}_{1_1}=$ \textit{male}, $\mathbf{z}_{1_2}=$ \textit{female} and $\mathbf{z}_{1_3}=$ \textit{other}, which are one-hot encoded as [1, 0, 0], [0, 1, 0] and [0, 0, 1] respectively. In this survey, we denotes the number of field as $m$ and the number of first-order features as $n$.

The co-existence of the original features is named high-order features (also known as feature interaction). More specifically, $\mathbf{X}^{(2)} = \{\langle \mathbf{x}_{i_p}, \mathbf{x}_{j_q}\rangle\}$ is considered second-order feature, $\mathbf{X}^{(3)} = \{\langle \mathbf{x}_{i_p}, \mathbf{x}_{j_q}, \mathbf{x}_{k_r}\rangle\}$ is considered third-order feature.

As we stated in Section \ref{sec:intro}, the two essential issues regarding the CTR feature representation learning are (i) which features we should represent and (ii) how we should represent these selected features. We will formulate them in detail in Section \ref{sec:form_which} and \ref{sec:form_how} respectively. Finally, we represent the overall learning criteria in Section \ref{sec:form_overall}.

\subsection{Formulation of the Which}
\label{sec:form_which}

The first issue regarding which feature to represent can be viewed as learning a binary tensor $\mathcal{G}^{(t)} \in \{0, 1\}^{C^t_n}$ for $\mathit{t}$-th order features. If certain features are chosen to be represented, the corresponding value in the tensor $\mathcal{G}^{(t)}_i$ becomes one. In contrast, if we do not represent certain features, the corresponding value in tensor $\mathcal{G}^{(t)}_i$ is zero. All these binary tensors are concatenated to represent the selection of features of all orders as:

\begin{equation}
    \mathcal{G} = [\mathcal{G}^{(1)}, \mathcal{G}^{(2)}, \dots ],
\end{equation}
which is served as input for the CTR prediction.

However, such binary tensor $\mathcal{G}^{(t)}$ with any $t \ge 2$ is extremely large. To efficiently explore such ample space, previous methods reduce the selection granularity to field level, which can be written as learning a binary tensor $\mathcal{\hat{G}}^{(t)} \in \{0, 1\}^{C^t_m}$.

\subsection{Formulation of the How}
\label{sec:form_how}

As for the way of representing selected features, it can be formulated as learning a transformation function $f^{(t)}()$ for $t$-thy order feature.

For first-order features, where $t=1$, it has been a common practice~\cite{FM,DeepFM} to transform them into dense vectors throughout the embedding layer. Such transformation can be formulated as follows:
\begin{equation}
    \mathbf{e}_{i_p} = f^{(1)}(\mathbf{x}_{i_p}) = \mathbf{E} \times \mathbf{x}_{i_p}.
\end{equation}
And we write all the input embedding as $\{\mathbf{e}_{i_p}\}$. Various methods~\cite{AutoDim,AutoEmb,OptEmbed} have been proposed to explore the suitable dimension of the embedding layer $\mathbf{E}$ instead of giving a pre-defined number.

For higher-order features, the representation methods are more diverse~\cite{OptInter}. Here we take second-order features $\mathbf{X}^{(2)} = \{\langle \mathbf{x}_{i_p}, \mathbf{x}_{j_q} \rangle\}$ as an example. The majority of the previous work adopt certain aggregation functions to integrate both first-order features. It can be formulated as:

\begin{equation}
    \mathbf{e}_{\langle i_p, j_q \rangle} = f^{(2)}(\langle \mathbf{x}_{i_p}, \mathbf{x}_{j_q} \rangle) = g(\mathbf{x}_{i_p}, \mathbf{x}_{j_q}),
\end{equation}
where the aggregation function $g()$ can be inner product~\cite{FM}, outer product~\cite{PNN} or cross layer~\cite{DCN}.

Meanwhile, certain method~\cite{OptInter} views the second-order features as cross-product feature, and represent them via additional embedding table $\mathbf{E}^{(2)}$ as well. This can be viewed as
\begin{equation}
    \mathbf{e}_{\langle i_p, j_q \rangle} = f^{(2)}(\langle \mathbf{x}_{i_p}, \mathbf{x}_{j_q} \rangle) = \mathbf{E}^{(2)} \times \langle \mathbf{x}_{i_p}, \mathbf{x}_{j_q} \rangle.
\end{equation}

Here we write all-order feature representations as 
\begin{equation}
    \mathcal{E} = [\{\mathbf{e}_{i_p}\}, \{\langle \mathbf{e}_{i_p}, \mathbf{e}_{j_q} \rangle \}, \dots].
\end{equation}

\subsection{Overall Training Criteria}
\label{sec:form_overall}

In summary, the feature representation learning target can be viewed as follows:
\begin{equation}
\label{eq:final_predict}
    \hat{y} = \mathcal{F}(\mathcal{G}, \mathcal{E})
\end{equation}

Here, the classifier $\mathcal{F}(\cdot)$ takes both all binary features selection mask of each order and representation embedding as input and outputs the prediction $\hat{y}$. 

As for the classifier, $\mathcal{F(\cdot)}$, various designs have been proposed recently~\cite{DCN}. These methods can mainly be categorized into three classes: shallow~\cite{FM}, deep~\cite{PNN} or hybrid~\cite{DeepFM,DCN}. However, the design of these classifiers is orthogonal to the representation learning of features. 

The final prediction is calculated based on Equation \ref{eq:final_predict}. The cross-entropy loss (i.e. log-loss) is adopted to measure the difference between prediction and label for each sample:
\begin{equation}
\label{eq:logloss}
    \text{CE} (y,\hat{y}) = y\log(\hat{y}) + (1-y)\log(1-\hat{y}),
\end{equation}
where $y$ is the ground truth of user clicks. We summarize the final accuracy loss as follows:
\begin{equation}
\label{eq:summarize}
    \mathcal{L}_{\text{CE}}(\mathcal{D}) = -\frac{1}{ |\mathcal{D}|} \sum_{(\mathbf{x}, y)\in\mathcal{D}} \text{CE}(y, \mathcal{F}(\mathcal{G}, \mathcal{E})),
\end{equation}
where $\mathcal{D}$ is the training dataset.

\section{Which Feature to Represent}
\label{sec:which}

\begin{figure}[!htbp]
    \centering
    \includegraphics[width=0.5\textwidth]{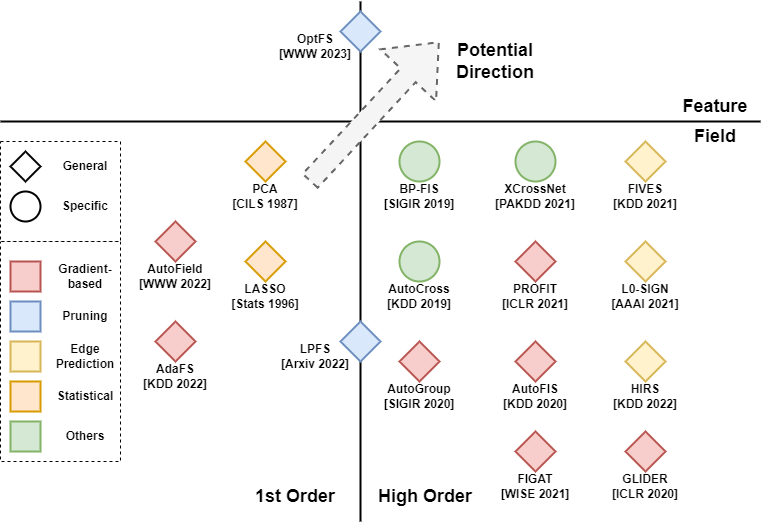}
    \caption{Summary and potential direction regarding which feature to represent. Here different colors indicates different approaches utilized by corresponding method. Different shape indicates the backbone model the methods applied.}
    \label{fig:which}
\end{figure}

Selecting informative features to represent is essential for accurate prediction in the prediction model. Multiple methods have been developed in the CTR prediction to efficiently boost performance. Moreover, the co-existence of features has also played an important role in CTR prediction~\cite{OptInter}. Therefore, we divide these methods into two categories: those who select the first-order feature and those which select the second or high-order feature, which will be described in detail. The summary table is shown in Table \ref{tab:which}. An illustration figure is also provided in Figure \ref{fig:which}, indicating both recent progresses and potential direction in this issue.

\begin{table*}
    \centering
    \begin{tabular}{l|cccc}
        \toprule
        Methods & Order & Granularity & Backbone Model & Approach \\
        \midrule
        PCA~\shortcite{PCA} & first & Field & General & Statistical \\
        LASSO~\shortcite{LASSO} & first & Field & General & Statistical \\
        LPFS~\shortcite{LPFS} & first\&high & Field & General & Pruning \\
        AutoField~\shortcite{AutoField} & first & Field & General & Gradient-based \\
        AdaFS~\shortcite{AdaFS} & first & Sample & General & Gradient-based \\
        OptFS~\shortcite{OptFS} & first\&high & Feature & General & Pruning \\
        \midrule
        BP-FIS~\shortcite{BP-FIS,BP-FIS-2} & first\&second & Field & FM~\shortcite{FM} & Bayesian Variation Selection\\
        AutoCross~\shortcite{AutoCross} & high & Field & Cross~\shortcite{DCN} & Beam Search \\
        GLIDER~\shortcite{GLIDER} & high & Field & General & Gradient-based \\
        AutoFIS~\shortcite{AutoFIS} & high & Field & General & Gradient-based \\
        FIGAT~\shortcite{FIGAT} & high & Field & General & Gradient-based \\
        AutoGroup~\shortcite{AutoGroup} & high & Field & General & Gradient-based \\
        PROFIT~\shortcite{PROFIT} & high & Field & General & Gradient-based \\
        XCrossNet~\shortcite{XCrossNet} & high & Field & Cross~\shortcite{DCN} & Gradient-based \\
        L0-SIGN~\shortcite{L0-SIGN} & second & Field & General & Edge Prediction \\
        HIRS~\shortcite{HIRS} & high & Field & General & Edge Prediction \\
        FIVES~\shortcite{FIVES} & high & Field & General & Edge Prediction \\
        \bottomrule
    \end{tabular}
    \caption{Summary of Which Feature to Represent. 
    \textit{General} means backbone model can be any prediction model.
    }
    \label{tab:which}
\end{table*}

\subsection{First-Order Feature}

The selection of the first-order feature can be formulated as learning a binary tensor $\mathcal{G}^{(1)} \in \{0,1\}^n$, which can be a large space in online CTR prediction. To reduce the selection space, various methods propose selecting informative feature fields instead of each field. These methods reduce the binary selection tensor $\mathcal{G}^{(1)}$ from $\{0,1\}^n$ to $\{0,1\}^m$, thus making the space feasible. 

Statistical-based methods~\cite{LASSO,PCA} exploit the statistical metrics of different feature fields and conduct feature field selection. PCA~\cite{PCA} projects each data point to a few components to obtain low-dimensional data while preserving the data variation as much as possible. It ranks the feature fields by the sum of absolute coefficients in all components and pick up a few feature fields with the highest value. LASSO~\cite{LASSO} select feature field by enforcing the sum of squared regression coefficients to be less than a predefined value and finally sets certain coefficients to zero, thus selecting the informative ones. 

Inspired by recent development of neural architecture search(NAS)~\cite{DARTS,NAO}, NAS-based methods have been proposed~\cite{AutoField,LPFS,AdaFS} to select first order features for CTR models. AutoField~\cite{AutoField} introduce a learnable value for each feature field, indicating the keep or drop of this field. It determines the value for each feature field automatically given the gradient-based NAS result~\cite{DARTS} and alternatively update the learnable values and network parameters. AdaFS~\cite{AdaFS} follows the same setting like AutoField~\cite{AutoField}. But it proposes a novel controller network to decide whether to select each feature fields according to the input sample, which fits the dynamic recommendation scenario better. LPFS~\cite{LPFS} replaces the gradient-based selector with a smoothed-$\mathit{L}_0$ optimization, which can be used to efficiently select second or high-order feature fields.

However, reducing the selection granularity from feature to field is too coarse to determine subtle features. OptFS~\cite{OptFS} efficiently explore the feature-level selection space $\mathcal{G}^{(1)} \in \{0,1\}^n$ and determine which feature to represent via a learning-by-continuation training scheme. It designs a decomposition between first and high-order feature selection to make accurate predictions.

\subsection{High-Order Feature}

The selection of $t$-th order feature can be formulated as learning binary tensor $\mathcal{G}^{(t)} \in \{0, 1\}^{C^t_n}$. It can be an extremely large selection space to explore. All previous methods~\cite{AutoFIS,L0-SIGN} adopt the field-level selection approximation to reduce the selection space from $\{0, 1\}^{C^t_n}$ to $\{0, 1\}^{C^t_m}$.

The selection of high-order features begins with algorithms targeting specific models.
BP-FIS~\cite{BP-FIS,BP-FIS-2} targets factorization machine~\cite{FM} views the selection of second-order features as learning personalized sparse factorization machines. It proposes Bayesian variable selection methods to reduce the number of second-order features. 
AutoCross~\cite{AutoCross} explicitly targets high-order cross features~\cite{DCN} and searches for useful ones in a tree-based search space by an efficient beam search method.

As for the general models, AutoFIS~\cite{AutoFIS} propose to learn the high-order features by adding an attention gate to each possible one, rather than enumerating all possible combinations. 
Based on that, AutoGroup~\cite{AutoGroup} considers the selection of high-order features as a structural optimization problem and alternatively updates the structural parameters and network parameters (e.g. the embedding layer and predictors) by gradient-based neural architecture search~\cite{DARTS}.
Following the setting of AutoFIS~\cite{AutoFIS}, PROFIT~\cite{PROFIT} propose to decompose the selection of $k$-th order feature fields via a symmetric decomposition. It also adopts gradient-based NAS to solve this problem like AutoGroup~\cite{AutoGroup}. 
GLIDER~\cite{GLIDER} propose to interpret feature interactions from a source model and explicitly encode these interactions in a target model, without assuming the model structure. 
FIGAT~\cite{FIGAT} proposes a novel model named Gated Attention Transformer, which can efficiently utilize the strong ability of the transformer while removing redundant high-order features. The vanilla attention mechanism provide interpretability and efficiency compared with self attention. 
XCrossNet~\cite{XCrossNet} proposes an efficient and interpretable way to learn high-order cross features containing dense and sparse first order features in an explicit manner. 

The selection of high-order features can also be viewed as an edge prediction problem when formulating each first-order feature as a node. L0-SIGN~\cite{L0-SIGN} first propose a formulation to select informative second-order features and solve the problem following the information bottleneck principle and statistical interaction theory. The authors further extend the definition of edge (connecting two nodes) to hyper-edge (connecting multiple nodes) and select high-order features via edge prediction~\cite{HIRS}. Also aiming to select high-order features under the graph settings, FIVES~\cite{FIVES} adopt the idea of layers in graph neural network. Once a $(t-1)$-th order feature is selected, its information is aggregated to a certain node in the GNN. A $t$-th order feature can be selected by predicting the edge between this node and other nodes representing first-order features.

\section{How to Represent Selected Features}
\label{sec:how}

How to represent selected features is also an essential issue in CTR representation learning.

\subsection{First-Order Feature}

Previous works~\cite{AutoDim,MDE,ATML} on representing first-order features can be categorized into two classes: Feature Hashing and Representation Dimension Selection. The former tends to hash features meeting the hardware requirement, while the latter aims to select a suitable representation dimension for each feature.

\subsubsection{Feature Hashing}

Hashing has been a classical way to improve model efficiency. In CTR prediction, feature hashing~\cite{AutoDis,CE,DHE} has been adopted to represent first-order features. 
AutoDis~\cite{AutoDis} focuses on representing the numerical features. It introduces meta-embeddings for each numerical field to represent the global knowledge of that field with a manageable number of parameters. Soft discretization is adopted to capture the correlation between meta-embeddings and numerical features, and representations are learnt through an aggregation function. 

Other works~\cite{CE,DHE} focus on hashing the categorical features. 
In Compositional Embeddings~\cite{CE}, the representation of each feature is represented as the combination of multi-embeddings determined by the complementary partitions. Hence, multiple smaller embedding tables instead of one large one are stored to improve efficiency while preserving the uniqueness of each feature.
DHE~\cite{DHE} first encodes the feature value to a unique identifier vector via multiple hashing functions and transformations and then applies a DNN to convert the identifier vector to a representation. The encoding module is deterministic, non-learnable, and free of storage, while the embedding network is updated during the training time to learn representation generation.

\subsubsection{Representation Dimension Selection}

\begin{table*}
    \centering
    \begin{tabular}{l|ccccc}
        \toprule
        Methods & Granularity & Flexible & Continuous & Removable & Approach \\
        \midrule
        MDE~\shortcite{MDE} & Field & \XSolidBrush & \Checkmark & \XSolidBrush & Heuristic \\
        EMDE~\shortcite{EMDE} & Field & \XSolidBrush  & \Checkmark & \XSolidBrush & Heuristic\\
        \midrule
        ESAPN~\shortcite{ESAPN} & User\&Item & \XSolidBrush & \Checkmark & \XSolidBrush & Reinforcement Learning \\
        NIS~\shortcite{NIS} & Feature Group & \XSolidBrush & \Checkmark & \XSolidBrush & Reinforcement Learning \\
        DNIS~\shortcite{DNIS} & Feature Group & \Checkmark & \Checkmark & \XSolidBrush & Gradient-based \\
        AutoEmb~\shortcite{AutoEmb} & User\&Item & \XSolidBrush & \Checkmark & \XSolidBrush & Gradient-based \\
        AutoDim~\shortcite{AutoDim} & Field & \XSolidBrush & \Checkmark & \XSolidBrush & Gradient-based \\
        \midrule
        ATML~\shortcite{ATML} & Feature Group & \Checkmark & \Checkmark & \XSolidBrush & Pruning \\
        RULE~\shortcite{RULE} & Item Group & \XSolidBrush & \XSolidBrush & \XSolidBrush & Evolutionary Search \\
        PEP~\shortcite{PEP} & Feature & \Checkmark & \XSolidBrush & \Checkmark & Pruning \\
        Autosrh~\shortcite{Autosrh} & Feature & \Checkmark & \XSolidBrush & \Checkmark & Pruning \& Gradient-based\\
        i-Razor~\shortcite{iRazor} & Field & \XSolidBrush & \Checkmark & \Checkmark & Gradient-based \\
        OptEmbed~\shortcite{OptEmbed} & Feature & \Checkmark & \Checkmark & \Checkmark & Pruning \& Evolutionary Search \\
        \bottomrule
    \end{tabular}
    \caption{Summary of Dimension Search Method. 
    Here \textit{Approach} represents the approach to solve the formulation.
    \textit{Flexible} means that pre-defined masks are not required.
    \textit{Continuous} means that the selected representation is continuous.
    \textit{Removable} means that certain representations can be completely removed.
    }
    \label{tab:dimension}
\end{table*}

Selecting suitable dimensions to represent first-order features has attracted many works. It has been pointed out that the over-parameterizing features with smaller feature cardinality may induce overfitting, and features with larger cardinality need larger dimensions to convey fruitful information~\cite{OptEmbed}. 

The most straightforward way is to incorporate expert experience on this topic heuristically. MDE~\cite{MDE} introduces a mixed-dimension embedding layer instead of a uniform dimension and heuristically assigns the dimension of each representation based on the feature's frequency. EMDE~\cite{EMDE} borrows the formulation of MDE~\cite{MDE} and proposes two variations of mixed embedding layer for matrix factorization.

Apart from heuristic approaches, other approaches~\cite{ESAPN,AutoDim,NIS} formulate the dimension selection as a neural architecture search problem. 
ESPAN~\cite{ESAPN} searches representation dimensions for both users and items dynamically based on popularity via an automated reinforcement learning agent. However, it requires pre-defined candidate dimension sets for both users and items.
Similarly, NIS~\cite{NIS} proposes to search suitable dimensions and vocabulary size via reinforcement learning-based neural architecture search~\cite{ENAS}. First, it ranks all features via frequency and groups neighbour features as one feature group as the basic search unit. Then, it selects from several pre-defined dimensions. 
DNIS~\cite{DNIS} borrows the same feature grouping mechanism as NIS~\cite{NIS} but relaxes the selection dimension in a more flexible space through continuous relaxation and differentiable optimization. 
AutoEmb~\cite{AutoEmb} proposes to use a controller network to dynamically select dimensions for both user and item based on their frequency under the recommender system settings.
AutoDim~\cite{AutoDim} extends the AutoEmb~\cite{AutoEmb}'s formulation to arbitrary fields. In addition, the Gumbel-Softmax~\cite{Gumbel-Softmax} trick is utilized instead of a controller network to determine the dimension of representations.

Except for viewing dimension selection as a neural architecture search problem, several research works~\cite{HAM,PEP} also formulate it as a pruning problem.
HAM~\cite{HAM} proposes to structurally prune the dimension of representation via a hard auxiliary mask at the field level. It first introduces an orthogonal regularity on embedding tables to reduce correlations within embedding columns and enhance representation capacity.
Inspired by the advancement of continuous sparsification~\cite{STR}, PEP~\cite{PEP} proposes to individually prune the redundant parameters of feature representation. 
Autosrh~\cite{Autosrh} adopts both pruning and gradient-based NAS approach. It proposes to relax the selection dimension like DNIS~\cite{DNIS} via gradient-based NAS and achieves a specific compression ratio with pruning parameters.

If we consider representation dimensions equaling zero as one of the possible dimensions, the selection of first-order features can be viewed as a special case of the dimension selection problem. 
i-Razor~\cite{iRazor} introduces dimensions equaling zero as one of the search candidates and proposes a differentiable neural architecture search mechanisms to simultaneously consider the "how" and "which" over the first-order features.
OptEmbed~\cite{OptEmbed} proposes to consider these two problems uniformly jointly. It efficiently trains a supernet with informative first-order features and embedding parameters simultaneously. Then, a one-shot evolutionary search method based on that supernet is conducted to produce an optimal embedding table.

\begin{table*}[!htbp]
    \centering
    \begin{tabular}{l|cc}
        \toprule
        Methods & Aggregation Func. Candidates & Approach \\
        \midrule
        AutoCTR~\shortcite{AutoCTR} & \{SLP, FM~\shortcite{FM}, IP\} & Evolutionary Search \\
        AutoRec~\cite{AutoRec} & \{SLP, Concate, Self-attention, IP, Random\} & \{Random, Grid, BO\}\\
        AutoFeature~\shortcite{AutoFeature} & \{Add, OP, Concate, GP, Null\} & Naive Bayes \\
        AutoPI~\shortcite{AutoPI} & \{Skip, SE layer~\shortcite{FiBiNet}, Self-attention, SLP, 1d Conv\} & Gradient-based \\
        NAS-CTR~\shortcite{NAS-CTR} & \{MLP, Cross~\shortcite{DCN}, FM~\shortcite{FM}, EW\} & Proximal Algorithm\\
        AutoIAS~\shortcite{AutoIAS} & \{OP, Concate, Add, Max\} & Reinforcement Learning \\
        AIM~\shortcite{AIM} & \{IP, OP, KP\} & Pruning \\
        OptInter~\shortcite{OptInter} & \{Embed, OP, Null\} & Gradient-based \\
        \bottomrule
    \end{tabular}
    \caption{Summary of High Order Feature Representation Method. 
    Here \textit{Approach} represents the approach to solve the corresponding formulation.
    \textit{IP}, \textit{OP}, \textit{SLP}, \textit{MLP}, \textit{KP}, \textit{GP}, \textit{Skip} stands for inner product, outer product, single-layer perceptron, multiple-layer perceptron, kernel product, generalized product and skip connect.
    \textit{EW} stands for multiple element-wise interactions, including add, average, inner product, max and min.
    \textit{Embed} means using an explicit embedding table to represent feature.
    \textit{Concate} means concatenation.
    \textit{Null} means that the representation is a zero vector.
    \textit{Grid} stands for grid search.
    \textit{BO} stands for Bayesian Optimization.
    }
    \label{tab:integration}
\end{table*}

The dimension selection problem over the first-order features may also align with other specific issues except for boosting the model performance. ATML~\cite{ATML} targets the warm-start-based applications in industrial scenarios and proposes an Adaptively-Masked Twins-based Layer behind the embedding layer to mask the undesired dimensions of each feature representation in a continuous manner. However, it also groups features via their frequency like previous works~\cite{NIS,DNIS}. Another work, RULE~\cite{RULE}, focuses on the easy deployment of a well-trained model to arbitrary device-specific memory constraints without retraining. It proposes an elastic embedding as concatenating a set of embedding blocks. An estimator-based evolutionary search function is designed to choose among the elastic embedding. 

\subsection{High-Order Feature}

As for high-order features, the key issue relies on how to generate their representation. 

Most of the previous work~\cite{AutoCross,AutoFeature,AutoPI} represents high-order features as the output of an aggregation function or network that takes multiple. 
AutoCTR~\cite{AutoCTR} proposes a learning-to-rank guided evolutionary search process and selects certain combinations from three different integration functions.
AutoRec~\cite{AutoRec} 
AutoFeature~\cite{AutoFeature} limits the search space to easy functions, such as inner product, and designs an evolutionary search algorithm, NBTree, to determine the aggregation functions for high-order features.
Compared with AutoFeature~\cite{AutoFeature}, AutoPI~\cite{AutoPI} adopts a more general search space, and its candidate functions are extracted from representative hand-crafted works.
NAS-CTR~\cite{NAS-CTR} formulates the architecture search as a joint optimization problem with discrete constraints and proposes a differentiable search algorithm based on proximal iteration.
AutoIAS~\cite{AutoIAS} proposes an integrated search space to model high-order feature representations. Furthermore, a supernet trained with knowledge distillation is adopted to consistently predict the performance of each candidate architecture when exploring the search space via a policy network. 

Unlike previous work that generates the high-order feature representation as the output of a function, OptInter~\cite{OptInter} first proposes to represent high-order features explicitly through an embedding table. To reduce the large storage requirement due to explicit modelling, it utilizes a gradient-based neural architecture search algorithm to select suitable representations for each field automatically. How to represent high-order feature representations may be closely related to how to represent first-order ones, as the former is usually calculated based on the latter.
AIM~\cite{AIM} also incorporates these two issues in a unified framework and utilizes gradient-based method techniques to generate sparse outcomes.

\section{Summary and Future Perspectives}
\label{sec:summary}

Over the past several years, feature representation learning has continued to be an inspiring and exciting topic in the click-through rate prediction domain. Both industrial and academic researchers have conducted many studies in this domain. In this review, we propose a taxonomy shown in Figure \ref{fig:taxonomy} that separates previous works following two major issues: (i) which feature to represent and (ii) how to represent selected features. These two issues are also unified into the same format in Section \ref{sec:form}. Detailed explanations of how different methods approach these two issues can be found in Section \ref{sec:which} and \ref{sec:how}, respectively. Despite the significant progress in feature representation learning in recent years, we note that significant challenges and open issues still exist in this domain. We summarize the potential perspectives as follows:

\paragraph{Fine-grained Feature Selection} 
Currently, the feature selection methods have to compromise the massive possible search space and reduce their granularity to the field level, greatly reducing the number of possible combinations during the process. This phenomenon is particularly severe when modelling higher-order features. Major advances in this domain are summarized in Figure \ref{fig:which} as well. In the near future, we would be delighted to see more works exploring the selection of high-order features in a more fine-grained manner.

\paragraph{Integrated Consideration of Different Order Features}
Most previous works only focus on selecting meaningful features in a certain order. They either select the first-order features and keep all the high-order ones or select the high-order features while fixing the first-order features. The former fails to filter useless high-order features, leading to higher computation costs and degraded model performance. The latter identifies useful high-order features from all available first-order features, resulting in redundant first-order features. It could be a promising topic to consider the selection of different order features jointly.

\paragraph{Integration of Both Issues}
It appears that most works focus on answering one issue: either the "which" or the "how". However, some researchers tries to unify both issues at different level. Autosrh~\cite{Autosrh}, i-Razor~\cite{iRazor}, PEP~\cite{PEP} and OptEmbed~\cite{OptEmbed} jointly consider the "which" and "how" for first-order features. AutoFeature~\cite{AutoFeature} and OptInter~\cite{OptInter} include null representation as one of the possible ways to represent high-order features, which equals to drop it. The "which" and "how" are closely related issues, as the way to represent features may be influenced by the selected features and vice versa. It is of great potential to integrate both issues.

\clearpage

\bibliographystyle{named}
\bibliography{ijcai23}

\end{document}